\documentclass[12pt]{article}

\begin{document}
\begin{titlepage}
\begin{center}

\vskip 1cm


{\large \bf {Yukawa Textures, New Physics and Nondecoupling}}

\vskip 1cm

Gustavo C. Branco,\footnote{gbranco@ist.utl.pt} 
M. N. Rebelo  \footnote{rebelo@ist.utl.pt}
and J. I. Silva-Marcos \footnote{juca@cftp.ist.utl.pt}
\vskip 0.05in

{\em Departamento de F{\'\i}sica and Centro  de F{\'\i}sica
Te{\'o}rica de Part{\'\i}culas (CFTP),\\
Instituto Superior T\'{e}cnico, Av. Rovisco Pais, 1049-001
Lisboa, Portugal}

\end{center}

\vskip 3cm

\begin{abstract}
We point out that New Physics can play an important r\^ ole
in rescuing some of the Yukawa texture zero ans\" atze which would
otherwise be eliminated by the recent, more precise
measurements of $V_{CKM}$. As an example, a
detailed analysis of
a four texture zero ansatz is presented, showing
how the presence of an isosinglet vector-like quark which
mixes with standard quarks, can render viable this
Yukawa texture.
The crucial point is the nondecoupling of the effects 
of the isosinglet quark, even for arbitrary large values of its mass. 
\end{abstract}

\end{titlepage}

\newpage
\section{Introduction}

The increasingly higher precision  in the determination of 
the elements of the fermion mixing matrices, both in the quark and lepton
sectors is clearly one of the most significant recent developments 
in particle physics and provides a great challenge to flavour models.

In most of the attempts at understanding the observed pattern of 
fermion masses and mixing, one assumes the existence of family 
symmetries either abelian or non-abelian, leading to special 
flavour structures in the Yukawa matrices, often involving texture 
zeros and/or a Froggatt-Nielsen type \cite{Froggatt:1978nt}
power structure of the matrix 
elements, in terms of a small expansion parameter. 
In the search for the allowed texture zeros, one
may take a bottom-up approach, where one uses the input data on
fermion masses and mixing to derive the Yukawa textures which are allowed
by experiment. Some years ago, 
Ramond, Roberts and Ross (RRR) \cite{Ramond:1993kv}, in
a pioneering work, followed this bottom-up approach and made a 
systematic search for allowed quark Yukawa structures. Assuming
symmetric or Hermitian Yukawa matrices and using the experimental
data available at the time, RRR found a total of five possible
solutions in a survey of all six and five texture-zero ans\" atze. 
Meanwhile, with the impressive improvement in the experimental 
determination of the $V_{CKM}$ matrix, all the texture-zero 
structures found in 
\cite{Ramond:1993kv} have great difficulty in reproducing the data. One of
the greatest challenges to these models arises from the 
precise determination of the rephasing invariant angle 
$\beta \equiv \arg \left( - V_{cd} V^{\ast}_{cb} V^{\ast}_{td} V_{tb}\right)$.
Indeed, it has been pointed out \cite{Branco:2004ya}
that in a large class of texture zero models which include 
all those considered by RRR, one cannot have a
sufficiently large value of $\sin (2 \beta) $, to conform to
the present experimental value 
$ \sin (2 \beta) = 0.687 \pm 0.032 $ \cite{Yao:2006px}

Another important constraint arises from the experimental
value of $B_{d}^{0}$--$\bar{B}_{d}^{0}$ mixing, combined
with the recent measurement of $B_{s}^{0}$--$\bar{B}_{s}^{0}$
mixing by D0 \cite{Abazov:2006dm} and CDF \cite{unknown:2006mq},
which leads to the extraction of the ratio 
$\left| V_{td}\right| / \left| V_{ts}\right|$
with relatively small errors. Very recently, the
rephasing invariant phase 
$\gamma \equiv \arg \left( - V_{ud} V^{\ast}_{ub} V^{\ast}_{cd} V_{cb}\right)$
has been measured by Belle \cite{Abe:2004gu}, \cite{Abe:2005ct}, 
\cite{Poluektov:2006ia}, and BaBar  \cite{Aubert:2005yj}
leading to the value $\gamma = \left(63^{+15}_{-12} \right)^{\circ} $ 
\cite{Yao:2006px}.  
In spite of the large experimental errors, the measurement of
$\gamma$ is of crucial importance due to the fact 
that its extraction from input data is 
essentially not affected by the presence of New Physics (NP)
contributions to $B_{d}^{0}$--$\bar{B}_{d}^{0}$  and
$B_{s}^{0}$--$\bar{B}_{s}^{0}$ mixings.

In the study of the impact of NP on the test of Yukawa textures, one
has to specify what are the assumptions on the nature of NP. In most
of the NP scenarios considered in the literature, one usually
assumes that NP does not contribute significantly to
the tree level decays of strange and B-mesons. This implies that 
NP does not affect the extraction of $\left| V_{us}\right| $, 
 $\left| V_{cb}\right| $,  $\left| V_{ub}\right| $ and 
$\gamma$, from experimental data. 
With these four inputs, one can reconstruct the full
$V_{CKM}$ matrix and in particular the reference unitarity 
triangle \cite{Buras:2000dm}, \cite{Bona:2006ah}, \cite{Charles:2006yw}. 
However, we may assume that there may be significant 
contributions from NP to  $B_{d}^{0}$--$\bar{B}_{d}^{0}$ 
and  $B_{s}^{0}$--$\bar{B}_{s}^{0}$ mixing, thus affecting the 
extraction of   $\left| V_{td}\right| $,
 $\left| V_{ts}\right| $ as well as  $\beta$ from experiment. Apart from
these effects, we may assume in these scenarios that NP decouples
from low energy physics. This assumption holds in a large class of
models beyond the SM and in particular in  most of the
supersymmetric extensions of the Standard Model (SM).
Let us now consider the implications for the tests of 
Yukawa textures. If a particular texture predicts a value of $\beta$
and/or of  $\left| V_{td}\right| $ in disagreement with experiment,
it may be rescued by the presence of NP contributions to 
$B_{d}^{0}$--$\bar{B}_{d}^{0}$ mixing. However, this class of 
NP  cannot rescue a texture which predicts
a value  $\left| V_{ub}\right| / \left| V_{cb}\right| $ and/or
$\gamma$ in disagreement with experiment. This question is
specially relevant, due to the fact that some of the
most attractive Yukawa textures do have difficulty in
conforming to the measured values of 
$\left| V_{ub}\right| / \left| V_{cb}\right| $
and $\gamma$.

In this paper we will show that there is a class of NP which
can solve the above conflict and thus render viable those 
Yukawa textures. At this stage, it is worth recalling that
in most proposals for family symmetries which could shed some light
on the flavour puzzle, there is the underlying assumption 
that, even if the family symmetry is embedded  in a grand-unified 
theory, the heavy particles decouple and do not affect low 
energy physics, in particular the masses and mixing observable 
at low energies. In this paper, we will consider a scenario
with heavy fermions
where the decoupling does not occur and in
particular heavy fermions do 
affect the effective standard fermion mass matrices
at low energies.
We will show that the influence of heavy 
vector-like quarks is such that they may render viable some of
the texture zero structures which would otherwise be eliminated 
by the more precise data presently available on the $V_{CKM}$ matrix. 
The embedding of a family structure into a larger framework where
heavy fermions are included, has the following interesting feature.
Let us consider a flavour model based on 
$SU(3)_c \times SU(2)_L \times U(1) \times F$ where $F$ denotes a 
family symmetry responsible for the presence of a set of texture zeros
in the three by three quark mass matrices 
$M_u$ and $M_d$. This symmetry can be trivially embedded 
into a larger framework with isosinglet vector-like heavy quarks
$Q$  by assuming that the SM fields keep their transformation 
properties under $F$ and allowing for  $F$ to be softly broken by 
$SU(3)_c \times SU(2)_L \times U(1)$ invariant
mass terms connecting $Q$ to standard quarks.
The striking feature of the example we will consider, 
with one singlet down-type vector-like quark, is that its effect 
in the low energy standard fermion masses and mixing 
can be sizeable even in the limit where heavy quark masses are  
very large and deviations from unitarity of the  $V_{CKM}$ matrix are
arbitrarily suppressed.

The paper is organized as follows. In the next section, we
analyse in detail a four texture zero Hermitian ansatz and
illustrate its difficulties in accommodating the 
present data on $V_{CKM}$. In section 3, we present an 
example of nondecoupling of NP and analyse in an 
analytical qualitative way how the presence of vector-like 
quarks can render the four texture zero ansatz compatible 
with our present knowledge on  $V_{CKM}$. In section 4, we 
provide an explicit example which is solved numerically, 
confirming our analysis of section 3. Finally, we present 
our conclusions in section 5.

\section{A Four Texture Zero Hermitian Ansatz}

Several Hermitian ans\" atze with texture zeros 
have been studied in the literature. These ans\" atze 
lead in general to predictions which usually consist of simple
relations for the mixing angles expressed in terms of quark mass 
ratios. It is worth emphasizing that Hermiticity is as important
as the existence of texture zeros, in order to obtain
predictive ans\" atze. Indeed, it has been 
shown  \cite{Branco:1988iq} that if one drops the 
requirement of Hermiticity,  most
of the texture-zero ans\" atze can be obtained, starting from
arbitrary quark mass matrices $M_u$, $M_d$, by simply making 
weak-basis transformations. This shows that without the
requirement of Hermiticity those texture zeros have no physical 
implications.

For definiteness, we consider a specially 
interesting four zero ansatz  which has 
been analysed  in detail  in the literature
\cite{Fritzsch:1999ee}, \cite{Roberts:2001zy}. 
The quark mass matrices $M_{u}$, $M_{d}$
are assumed to have the form:
\begin{equation}
M_{u}=\lambda _{u}\ \ K_{u}^{\dagger }\ \ \left[
\begin{array}{ccc}
0 & a_{u} & 0 \\
a_{u} & b_{u} & c_{u} \\
0 & c_{u} & 1-b_{u}
\end{array}
\right] \ K_{u}\quad ;\quad M_{d}=\lambda _{d}\ \ \left[
\begin{array}{ccc}
0 & a_{d} & 0 \\
a_{d} & b_{d} & c_{d} \\
0 & c_{d} & 1-b_{d}
\end{array}
\right]  
\label{massas}
\end{equation}
where $K_{u}=diag(e^{i\phi _{1}},1,$ $e^{i\phi _{3}})$ 
and all other parameters are real.

It is clear from Eq.~(\ref{massas}) that the trace of each matrix 
was factored out so that one has, by construction:
\begin{equation}
\mathrm{Tr}(M_{u})=\lambda _{u}\quad ;\quad \mathrm{Tr}(M_{d})=\lambda _{d}
\label{trace}
\end{equation}
The convention of phases adopted in $K_{u}$ corresponds to the factoring
out of all phases in $M_u$ and $M_d$ and the elimination of the
maximum number of non-physical ones. It is clear that no 
non-factorizable phases remain in this ansatz due to the existence of one
zero off-diagonal entry in both $M_u$ and $M_d$. 
It was shown in Ref. \cite{Branco:2004ya}
that the absence of nonfactorizable phases leads to important restrictions 
on $\sin 2 \beta$.

The presence of several zeros in this Hermitian ansatz renders
the analytical diagonalization of the mass matrices quite simple.
The column vectors of the unitary matrices $U^u$, $U^d$ which 
diagonalize $M_u$, $M_d$ can be determined, in each case, via the vector
product of the first and third rows, 
with the inclusion of the mass eigenvalues.
Each one of the three columns can be expressed as:
\begin{equation}
(-m_i^\prime ,\  a,\  0) \times (0,\  c, \  1-b-m_i^\prime) 
\qquad \frac{1}{N_i}
\label{eiv}
\end{equation}
where we have omitted the $u$, $d$, sub-indices,
$N_i$ is a normalization factor and $m_i^\prime$  
denotes the ith mass
eigenvalue divided by the sum of the three mass eigenvalues. 
The mass eigenvalues do not depend on $K_u$ and thus both in the up and 
and down quark sectors, the four parameters $ a, b, c, \lambda$, 
can be expressed in terms of quark masses, leaving only one free 
parameter in each sector, which we may chose to be $b_{(u,d)}$.
This allows one to write each of the unitary matrices
$U^u$, $U^d$  in terms of quark mass ratios, a single free parameter
and the phases $\phi_1$, $\phi_3$ which have been factored out.

In this way one obtains the well known texture zero relations 
\cite{Hall:1993ni}, \cite{Ramond:1993kv}, \cite{Barbieri:1997tu},
\cite{Roberts:2001zy} valid to leading order:
\begin{equation}
\left| \frac{V_{ub}}{V_{cb}} \right| = \sqrt{\frac{m_u}{m_c}}
\qquad \left| \frac{V_{td}}{V_{ts}} \right| = \sqrt{\frac{m_d}{m_s}}
\qquad \left|V_{us} \right| = \left|
\sqrt{\frac{m_d}{m_s}}e^{i \phi_1} - \sqrt{\frac{m_u}{m_c}} \right| 
\label{eq4}
\end{equation}
which are verified by a wide class of models \cite{Hall:1993ni}.
Furthermore, as pointed out in the introduction, 
it has been shown  \cite{Branco:2004ya} that
texture zero ans\" atze with no non-factorizable phases,
such as this one,  cannot reach values 
of $\sin(2 \beta )$ as high as the present central value \cite{Yao:2006px}:
\begin{equation}
\sin(2 \beta ) =0.687 \pm 0.032
\label{s2b}
\end{equation}

It was already pointed out in  Ref. \cite{Roberts:2001zy}
that the relation obtained for $\left| \frac{V_{ub}}{V_{cb}} \right|$
was problematic, and strongly disfavoured this ansatz, due to
the smallness of the ratio   $\sqrt{\frac{m_u}{m_c}}$. At present the
constraint has become even more severe, since the new 
experimental average for  $\left| V_{ub} \right| $ went up significantly. 
The current experimental values for these two  $V_{CKM}$ entries are  
\cite{Yao:2006px}:
\begin{equation}
 \left| V_{ub}\right| =(4.31 \pm 0.30) \times 10^{-3}  \qquad
 \left| V_{cb}\right| =(41.6 \pm 0.6) \times 10^{-3} 
\label{ubcb}
\end{equation} 
whilst the values of the quark mass matrices
are taken as \cite{Yao:2006px}:
\begin{eqnarray}
m_u = 1.5 - 3.0 \ \mbox{(Mev)}, 
\ m_c = 1250 \pm 0.090 \ \mbox{(Mev)},\  m_t \simeq 300 \ \mbox {(Gev)} 
\nonumber \\
m_d = 3 - 7 \  \mbox{(Mev)},\  m_s = 95 \pm 25 \ \mbox{(Mev)},  
\  m_b = 4.2 \pm 0.07  \mbox{(Gev)}, \\
m_u/m_d = 0.3 - 0.7 \qquad m_s/m_d = 17 - 22 \nonumber
\end{eqnarray}
The new theoretically clean and significantly improved
constraint on $\left| V_{td}\right| / \left| V_{ts}\right|$ 
is \cite{Yao:2006px}:
\begin{equation}
\left| V_{td}\right| / \left| V_{ts}\right| = 0.208^{+0.008}_{-0.006}.
\label{tdts}
\end{equation}
Taking into account the small experimental error, this results
deviates significantly from the value predicted by the ansatz,
$\left| V_{td} \right| /  \left| V_{ts} \right| = \sqrt{\frac{m_d}{m_s}}$
which leads to the range:
\begin{equation}
0.213 <\left| V_{td}\right| / \left| V_{ts}\right| < 0.243
\end{equation}
where we took into account the experimental constraint on  $m_s/m_d$
given above.

Next we briefly describe how the value of $\gamma$,
($\gamma \equiv \arg \left( - V_{ud} V^{\ast}_{ub} V^{\ast}_{cd} 
V_{cb}\right)$) for this ansatz can be derived analytically.
It is clear from Eq.~(\ref{massas}) that in our parametrization
$\arg  V_{ud} = \phi_1$, to an excellent approximation.. 
The phase $\phi_1$ is fixed by the experimental value of 
$\left| V_{us}\right|$, given by Eq.~(\ref{eq4}). Using the 
central values for the quark mass ratios:
\begin{equation}
\sqrt{\frac{m_d}{m_s}} = \sqrt{\frac{1}{20}} = 0.224;
\qquad \sqrt{\frac{m_u}{m_c}} = 0.042
\end{equation}
together with \cite{Yao:2006px} 
$\left| V_{us}\right| = 0.2257 \pm 0.0021$, one obtains
\begin{equation}
\phi_1 = -87^{\circ}
\end{equation}
Now, to leading order, one has $ V_{ub}/V_{cb} = - \sqrt{\frac{m_u}{m_c}}$,
implying $\arg ( V^{\ast}_{ub} V_{cb}) \simeq \pi $
and one obtains in good approximation:
\begin{equation}  
\gamma \simeq \arg \left( V_{ud} V^{\ast}_{cd} \right)
\end{equation}
Using the fact that in leading order in our parametrization
\begin{equation}
V_{cd} =  - \sqrt{\frac{m_d}{m_s}} +
\sqrt{\frac{m_u}{m_c}} e^{i \phi_1} 
\end{equation}
and taking central values for the quark mass ratios, one obtains 
\begin{equation}
\arg ( V^{\ast}_{cd}) = 169^{\circ} 
\end{equation}
Which finally leads to:
\begin{equation}
\gamma = \arg  V_{ud} + \arg V^{\ast}_{cd} = (-87^{\circ} +
169^{\circ}) = 82^{\circ}
\end{equation}
It is clear that in the framework of this ansatz, the value 
of $\gamma$ is very constrained, 
even allowing for one $\sigma$ deviations from central values of the
experimental parameters. At present, the current experimental value 
\cite{Yao:2006px}
$\gamma = \left(63^{+15}_{-12} \right)^{\circ} $ has large 
errors and therefore it is not possible to exclude the
ansatz only on the grounds of the $\gamma$ constraint. However, 
it is clear that the ansatz tends to give values for $\gamma$
larger than the central experimental value. 

Concerning $\beta$, it was shown in a previous work 
\cite{Branco:2004ya}
that only $\arg (V_{cd}) $ contributes significantly, so that
in this framework we have:
\begin{equation}
\beta \simeq  - \arg V_{cd}  \simeq 180^{\circ} - 169^{\circ}
= 11^{\circ}
\end{equation}

In summary, the four texture zero ansatz of  Eq.~(\ref{massas})
has serious difficulties in accommodating the recent, more
precise, experimental data on $V_{CKM}$. It is useful to separate 
these difficulties of the ansatz in two classes: \\

(i) The ansatz predicts too small a value for $\beta$ and
too large a value for  
$\left| V_{td}\right| / \left| V_{ts}\right| $. \\

(ii) The ansatz predicts too small a value for
 $\left| V_{ub}\right| / \left| V_{cb}\right| $
and too large a value for $\gamma$ \\

The important point we wish to emphasize is that, while
difficulties of class (i) can be avoided by assuming New Physics (NP) 
contributions to  $B_{d}^{0}$--$\bar{B}_{d}^{0}$ and 
$B_{s}^{0}$--$\bar{B}_{s}^{0}$ mixings, those of class  (ii)
remain a challenge to the ansatz even in the presence of
NP contributions to mixing. It is useful to parametrize NP
contributions to mixing in the following way:
\begin{equation}
M^{(q)}_{12} = ( M^{(q)}_{12} )^{SM} r^2_q e^{-2i\phi_q}
\qquad q=d,s
\end{equation}
The SM corresponds to $r_q =1$, $\phi_q =0$. In the presence of NP,
instead of $\beta$ one measures $\beta-\phi_d$ and 
$\Delta M_{B_q} = (\Delta M_{B_q})^{SM} r^2_q$. It is clear that 
even a small contribution of $\phi_d$ (i.e. $\phi_d = -11^{\circ}$)
together with a small deviation of $r_d/ r_s$ from unity can
rescue the ansatz from discrepancies
of class (i). On the contrary, 
the extraction of $\left| V_{ub}\right| / \left| V_{cb}\right| $
and $\gamma$ from experiment is unaffected by the presence
of NP in the mixing. At this stage it should be noted that in most
of the extensions of the SM, including the supersymetric ones, there are 
NP contributions to the mixing \cite{Branco:1995cj} -- \cite{Baek:1999jq}.
\\

An alternative way of checking that this interesting 
ansatz is in conflict with experiment is through the use of 
the following exact unitarity relation \cite{Botella:2002fr}:
\begin{equation}
\frac{\sin \beta }{\sin (\gamma +\beta )} =
\frac{\left| V_{ub}\right|}{ \left| V_{cb}\right|}
\frac{\left| V_{ud}\right|}
{\left| V_{cd}\right| }
\label{rel9}
\end{equation}
as well as another unitarity relation, that holds 
to an excellent approximation \cite{Botella:2002fr}:
\begin{equation}
\frac{\sin \gamma }{\sin (\gamma +\beta )} \simeq
\frac{\left| V_{td}\right| / \left| V_{ts}\right|}
{\left| V_{us}\right| }
\label{rel10} 
\end{equation}
Replacing in Eq.~(\ref{rel9})  the values obtained in this ansatz for
$\left| V_{ud}\right| \simeq 1$, $\left| V_{cd}\right| $
and the ratio $\left| V_{ub}\right| / \left| V_{cb}\right| $
we obtain the prediction
$\sin \beta / \sin (\gamma +\beta ) \simeq 0.19$. 
This is to be compared to the value computed with experimental central 
values   $[\sin \beta / \sin (\gamma +\beta)]_{\mbox{exp}} \simeq 0.37$.
Likewise for the relation given by Eq.~(\ref{rel10}), where
in this case the ansatz  predicts  
$\sin \gamma / \sin (\gamma +\beta ) \simeq 0.99$, while
the value computed with the experimental central values is 0.89.
We have also used the unitarity relations of Eqs.~(\ref{rel9}),
(\ref{rel10}) to verify the validity of our approximate analytical
evaluation of the elements of $V_{CKM}$, predicted by the ansatz.

Given the difficulties of this ansatz in conforming to the
experimental data, one may wonder whether the ansatz may 
be ``saved'' if implemented in a larger framework. 
In the next section, we show that this is indeed the case.
We describe a scenario
where the presence of NP can fully rescue the four 
texture zero ansatz by embedding it into a minimal extension 
of the SM with one additional down vectorial
isosinglet quark. This framework could result from a 
family symmetry of 
the Lagrangian leading to the existing texture zeros, which is
softly broken by mass terms involving the additional heavy quark.

\section{An example of nondecoupling}
Let us consider a model with only one $Q=-1/3$ isosinglet vector-like quark.
Vector-like quarks arise in a variety of extensions of the SM,
in particular, within the framework of grand-unified theories
based on $E_6$. Another motivation for introducing 
vector-like quarks arises if one requires spontaneous CP violation
\cite{Lee:1973iz} -- \cite{Branco:1985pf} in the context of 
supersymmetric extensions of the SM \cite{Branco:2000dq}, 
\cite{Hugonie:2003yu}.
Vector-like quarks are essential in order to 
generate a complex $V_{CKM}$ from vacuum phases \cite{Branco:2006pj}.
 
It can be easily shown that, without loss of generality, one may choose a 
weak basis where $M_u$,
the up quark mass matrix, is real and diagonal, and the down quark matrix 
${\cal M}_d$ can be cast in the form:
\begin{equation}
{\cal M}_d = \left(\begin{array}{ccccc}
  &  &   & | & 0  \\
  & m_d &  & | & 0  \\
 &  &  & | & 0  \\
- & - & - &  | & - \\
  & M_D & & | & H  \\  
\end{array}\right)
\label{qqqq}
\end{equation}
with $m_d$ a Hermitian $3 \times 3$ matrix, $ M_D $ a 
$1 \times 3$ matrix and $H$ a single entry. \\
The matrix ${\cal M}_d$ is diagonalized by the usual 
bi-unitary transformation:
\begin{equation}
U^{\dagger}_{L}{\cal M}_d U_{R} =  \left(\begin{array}{cc}
{\overline m} & 0 \\
0 & {\overline M} 
\end{array}\right)
\label{tttt}
\end{equation}
where ${\overline m} =$ diag $(m_d, m_s, m_b )$ and  
${\overline M} $ is the heavy quark mass. One can write $U_L$ in block form,
\begin{equation}
U_L  =  \left(\begin{array}{cc}
 K  & R \\
 S & T 
\end{array}\right)
\label{krst}
\end{equation}
where $K$ is the usual $ 3 \times 3$  $V_{CKM}$ matrix. $U_L$ is the matrix 
that diagonalizes  ${\cal M}_d {\cal M}^{\dagger}_d$, and the 
following relations can be readily derived \cite{Bento:1991ez}
in the limit $ M_D, H > > {\cal O} (m_d) $
\begin{eqnarray}
\overline{M} ^2 \simeq (M_D {M_D}^\dagger + H^2)  \equiv M^2 \\
\overline{m} ^2 \simeq K^{\dagger} m_{eff}  m^{\dagger}_{eff} K
\label{mem}
\end{eqnarray}
with 
\begin{equation}
 m_{eff}  m^{\dagger}_{eff} \simeq m_d {m_d}^\dagger -
\frac{(m_d {M_D}^\dagger M_D \  {m_d}^\dagger)} {M^2}
\label{eee}
\end{equation}
Note that $K$ is the mixing matrix connecting standard quarks
and has small deviations from unitarity given by
$K^{\dagger} K = 1 - S^{\dagger} S$, with:
\begin{equation}
 S \simeq - \frac{M_D m^\dagger_d K}{M^2} \left( 1 + 
\frac{\overline{m}^2} {M^2} \right)
\label{sss}
\end{equation}
At this stage, it should be noted that the mass terms $M_D$, $H$ 
are $SU(2) \times U(1)$ invariant and thus they can be much 
larger than the electroweak scale. If one makes the natural assumption 
that $M_D M^{\dagger}_D$ and $H^2$ are of the same order of magnitude
it is clear that in  Eq.~(\ref{eee}), the second term contributing to 
$ m_{eff}  m^{\dagger}_{eff}$, 
has a magnitude comparable to that of 
$m_d {m_d}^\dagger$. This is the crucial point which makes
it possible to rescue the four texture ansatz considered in
the previous section, through the introduction of a 
vector like isosinglet quark. Let us assume that there is a 
family symmetry which leads to the four texture zero ansatz, in the 
$3 \times 3$ quark mass matrices involving standard quarks.
The $SU(2) \times U(1)$ invariant mass terms  $M_D$, $H$ may break softly
the family symmetry. It is clear that the presence
of the second term contributing to $ m_{eff}  m^{\dagger}_{eff}$
in Eq.~(\ref{eee}),
does affect the predictions of the ansatz, allowing for it to be in
agreement with the present experimental data. In what follows, we 
explain how this is possible, first through a qualitative analysis 
and then in the next section  through an exact numerical example.

At this stage the following comment is in order. 
It is well known that in models with  isosinglet quarks
there are $Z$ mediated flavour changing neutral 
currents (ZFCNC) \cite{Nir:1990yq},\cite{Branco:1992wr}, with strength  
proportional to deviations of $3 \times 3$ unitarity of the 
$V_{CKM}$ matrix. From Eq.~(\ref{sss}) it is clear that deviations 
of unitarity are proportional to $\frac{m^2_d}{M^2}$ 
\cite{Branco:1986my}, \cite{Bento:1991ez}, and 
therefore are naturally suppressed. As a result, 
choosing both $M_D$ and $H$  much larger than $m_d$ 
strongly suppresses ZFCNC. This in turn implies that 
 for sufficiently large $M$ the extraction of $\beta$ and of 
$\left| V_{td}\right| / \left| V_{ts}\right|$ 
are not significantly changed from the one based on SM physics.

Let us consider the following structure for ${\cal M}_d $, with the
previous texture zero Hermitian ansatz embedded in the 
new four by four down mass matrix: 
\begin{equation}
{\cal M}_d = \left(\begin{array}{ccccc}
  &  &   & | & 0  \\
  & M_d &  & | & 0  \\
 &  &  & | & 0  \\
- & - & - &  | & - \\
  & M_D & & | & H  \\  
\end{array}\right) = 
\left(\begin{array}{cccc}
  0 & A & 0 & 0  \\
  A & B & C & 0  \\
  0 & C & D & 0  \\
  0 & f & g & H  \\  
\end{array}\right)
\label{qua}
\end{equation}
it is now possible to compute  $ m_{eff}  m^{\dagger}_{eff}$, 
to a good approximation, using Eq.~(\ref{eee}). We are now 
interested in combining this larger ansatz for ${\cal M}_d$ with  
$M_u$ following the previous pattern of Hermiticity and  
texture zeros.

It is known \cite{Branco:1999tw}, \cite{Roberts:2001zy}
that the present experimental data can be well
reproduced from the following Froggatt-Nielsen pattern for $m_{eff}$:
\begin{equation}
m_{eff} \sim m_b 
\left(\begin{array}{ccc}
  0 & \overline{\varepsilon}^3  &   \overline{\varepsilon}^4 \\
   \overline{\varepsilon}^3  &  \overline{\varepsilon}^2 &
   \overline{\varepsilon}^2    \\
   \overline{\varepsilon}^4 &  \overline{\varepsilon}^2 & 1   \\
\end{array}\right),
\label{exp}
\end{equation}
with $ \overline{\varepsilon} \simeq 0.2$ together with a similar
pattern for the up sector in terms of a smaller parameter $\varepsilon$
with a value close to 0.06.

The required structure for $m_{eff}  m^{\dagger}_{eff}$ is:
\begin{equation}
 m_{eff}  m^{\dagger}_{eff} \sim m_b^2  \left(\begin{array}{ccc}
\overline{\varepsilon}^6   & \overline{\varepsilon}^5  &   
\overline{\varepsilon}^4 \\
   \overline{\varepsilon}^5  &  \overline{\varepsilon}^4 &
   \overline{\varepsilon}^2    \\
   \overline{\varepsilon}^4 &  \overline{\varepsilon}^2 & 1   \\
\end{array}\right)
\label{mqd}
\end{equation}

Using Eqs.~(\ref{eee}), (\ref{qua} ) it can be verified  
that, starting from a Froggatt-Nielsen 
pattern for  $M_d$ given by:
\begin{equation}
|M_d| \sim m_b 
\left(\begin{array}{ccc}
  0 & \overline{\varepsilon}^3  & 0   \\
   \overline{\varepsilon}^3  &  \overline{\varepsilon}^2 &
   \overline{\varepsilon}^2    \\
   0 &  \overline{\varepsilon}^2 & 1   \\
\end{array}\right),
\label{eps}
\end{equation}
in the context of the  previous four texture ansatz,
implemented in the above extension of the SM, 
it is possible to obtain a matrix
$ m_{eff}  m^{\dagger}_{eff}$ with the structure given by 
Eq.~(\ref{mqd}), by assuming: 
$\frac{|fg|}{H^2} \sim  \overline{\varepsilon}$, in Eq.~(\ref{qua}).

This implies that the extension of the SM through the inclusion
of one down vector-like isosinglet quark can save the ansatz discussed in
the previous section. Since no additional quarks were introduced
in the up sector, 
the matrix $M_u$ remains unchanged. However, due to the different 
hierarchies of the quark masses in the up and down sector, 
the sensitivity of the $V_{CKM}$ matrix is much higher to changes 
in the down sector than to changes in the up sector.

\section{A numerical Example}
In this section we give an explicit example which illustrates
the above described framework. 

Let us consider the following mass matrices in Gev units:
 \begin{equation} 
{\cal M}_d = \left(\begin{array}{cccc}
  0 &  0.0258 & 0 & 0  \\
   0.0258 & 0.12  & 0.24 & 0  \\
  0 & 0.24 &  4.97 & 0  \\
  0 &  350 &  370i &  500 \\  
\end{array}\right); \qquad
M_{u}= \ K_{u}^{\dagger }\ \ \left[
\begin{array}{ccc}
0 &  0.056 & 0 \\
 0.056 & 1.3 &  2.8 \\
0 & 2.8 & 300
\end{array}
\right] \ K_{u}
\label{num}
\end{equation}
where $K_{u}=diag(e^{i\phi _{1}},1,$ $e^{i\phi _{3}})$ with
$\phi _{1} = -98.1^{\circ}$ and $\phi _{3} = 0.0^{\circ}$. It can
be readily verified that ${\cal M}_d$ and $M_{u}$ lead to the following 
masses and mixing:
\begin{eqnarray} 
(m_u,\ m_c,\ m_t)\  = \ ( 0.00246,\  1.28, \  300.0) \ \mbox{in GeV}; 
\nonumber \\
(m_d,\ m_s,\ m_b, \ M_4)\  = \ (0.0058,\ 0.0935,\ 4.3,\ 718.7) \ 
\mbox{in GeV}; \\
V_{CKM} (3 \times 4)  = \left(\begin{array}{cccc}
  0.9743 & 0.2252 & 0.0036 &   0.000013 \\
    0.2251&  0.9735 & 0.0410 & 0.00016  \\
   0.0084 & 0.0402 &   0.9991 &  0.0036 \\
\end{array}\right); \nonumber
\end{eqnarray}
The fourth column corresponds to matrix $R$ in Eq.~(\ref{krst}).
An interesting feature of this example is the extreme smallness 
of the deviations from unitarity of the $3 \times 3$
$V_{CKM}$ matrix as confirmed
by the smallness of all the entries in $R$. 

The corresponding values for $\sin (2\beta)$ and $\gamma$ are:
\begin{equation} 
\sin (2 \beta ) = 0.707, \qquad \gamma =  66.1^{\circ}
\end{equation} 
which are in good agreement with the present experimental bounds.
The ratio $\left| V_{ub}\right| / \left| V_{cb}\right|$ in this
example is equal to 0.088 and therefore is significantly
larger than $\sqrt {m_u /m_c} $. The fact that  $\left| V_{ub}\right| $
in our example does not exactly agree with the new experimental 
constraint of Eq.~(\ref{ubcb}) should not come as a surprise
since in our analysis we are constrained by unitarity. It can be 
easily checked that the present experimental central values
deviate from the unitarity relation of  Eq.~(\ref{rel9}),
although verifying 
Eq.~(\ref{rel10}). Note that, as emphasized in Ref.~\cite{Yao:2006px}
the new experimental average for $\left| V_{ub}\right| $
is somewhat above the range favoured by the measurement of 
$\sin (2 \beta)$.

The fact that physics at a high energy scale can save a low 
energy texture that, by itself, was already ruled out by experiment,  
is perhaps unexpected. More surprising even is the fact that
a similar effect could be obtained with a much heavier vectorial quark.
It was shown in a recent paper \cite{Higuchi:2006sx}
that down-type
vectorial isosinglet quarks can also play an important r\^ ole
in generating sufficient CP violation in models with
universal strength of Yukawa couplings
\cite{Branco:1990fj},\cite{Kalinowski:1990mz}.

\section{Conclusions}
We have studied the impact of New Physics on tests of Yukawa 
texture zero ans\" atze, emphasizing that the greatest
challenge for these textures arises from the measured values
of $\left| V_{ub}\right| / \left| V_{cb}\right|$ and the rephasing
invariant angle $\gamma$. This stems from the fact that
while the presence of New Physics contributions to
 $B_{d}^{0}$--$\bar{B}_{d}^{0}$  and/or 
$B_{s}^{0}$--$\bar{B}_{s}^{0}$ mixings can solve 
eventual discrepancies in the predictions for $\beta$, 
 $\left| V_{td}\right| $,  $\left| V_{ts}\right| $,
the extracted values of  $\left| V_{ub}\right| / \left| V_{cb}\right|$
and  $\gamma$ are unaffected by the presence of New Physics 
contributions to mixing. We then show that the presence
of New Physics which does not decouple at low energies
can save some of the most interesting ans\" atze which would
otherwise be in conflict with experiment. We illustrate these
effects through a specific four texture zero ansatz which is
studied in the context of a minimal extension of the SM 
with an isosinglet vector-like heavy quark which mixes with 
the standard quarks. We show that the presence of the heavy quark is
sufficient to render viable the ansatz which would otherwise 
be eliminated by the recent measurements of 
 $\left| V_{ub}\right| / \left| V_{cb}\right|$
and  $\gamma$. The crucial point is the nondecoupling 
of the effects of the heavy quark, even in the limit where its mass
is arbitrarily large. It is clear that analogous considerations
may in principle be applied to other texture zero ans\" atze 
which, without the input of New Physics would be ruled out by experiment.

\section*{Acknowledgements}
This work was partially supported by CERN and by Funda\c{c}\~{a}o para a 
Ci\^{e}ncia e a  Tecnologia (FCT, Portugal), through the projects 
POCTI/FNU/44409/2002, PDCT/FP/
FNU/50250/2003,
POCI/FP/63415/2005, and CFTP-FCT UNIT 777, which
are partially funded through POCTI (FEDER). The authors are grateful
for the warm hospitality of the CERN Physics Department (PH) Theory (TH)
where this work started. The work of G.C.B. was supported by the Alexander 
von Humboldt Foundation through a Humboldt Research Award. G.C.B. would
like to thank Andrzej J. Buras for the kind hospitality at 
Technische Universit\" at M\" unchen.

\end{document}